\documentclass{aastex}
\usepackage{emulateapj5,times,mathptm}
\usepackage{natbib,psfig}

\newcommand{\flux}{\hbox{erg~cm$^{-2}$~s$^{-1}$}}
\newcommand{\lumin}{{erg~s$^{-1}$}}

\newcommand{\msun}{\hbox{${M}_{\odot}$}}

\newcommand{\simgt}{\lower 2pt \hbox{$\, \buildrel {\scriptstyle >}\over {\scriptstyle\sim}\,$}}
\newcommand{\simlt}{\lower 2pt \hbox{$\, \buildrel {\scriptstyle <}\over {\scriptstyle\sim}\,$}}

\newcommand{\xmm}{{\emph{XMM-Newton}}}
\newcommand{\chandra}{{\emph{Chandra}}}
\newcommand{\einstein}{{\emph{Einstein}}}

\begin{document}


\submitted{to appear in the Astrophysical Journal Letters}
\title{Lower Mass Black Holes in the GOODS?
Off-nuclear X-ray Sources$^{1}$ }


\author{A.E.~Hornschemeier,$^{2,3}$ ~
D.M.~Alexander,$^{4,5}$
F.E.~Bauer,$^5$
W.N.~Brandt,$^5$
R.~Chary,$^6$ 
C.~Conselice,$^{7}$
N.A.~Grogin,$^{2}$,
A.M.~Koekemoer,$^{8}$ 
B.~Mobasher,$^{8}$
M.~Paolillo,$^{8}$,
S.~Ravindranath,$^{8}$
E.J.~Schreier$^{8}$
}

\altaffiltext{1}{Based on observations taken with the NASA/ESA Hubble
Space Telescope, which is operated by the Association of Universities for Research in
Astronomy, Inc.\ (AURA) under NASA contract NAS5--26555 and on observations collected
with the European Southern Observatory, Chile.}

\altaffiltext{2}{Johns Hopkins University, 3400 N. Charles Street, Baltimore, MD 21218}

\altaffiltext{3}{Chandra Fellow}

\altaffiltext{4}{Institute of Astronomy, Madingley Road, Cambridge, CB3 0HA, UK}

\altaffiltext{5}{The Pennsylvania State University, 525 Davey Laboratory, 
University Park, PA 16802}

\altaffiltext{6}{SIRTF Science Center, MS220-6, California Institute of Technology, Pasadena, CA 91125}

\altaffiltext{7}{California Institute of Technology, Pasadena, CA 91125}

\altaffiltext{8}{Space Telescope Science Institute, 3700 San Martin Drive, Baltimore, MD 21218}


\begin{abstract}

 We have identified a sample of 10 highly reliable off-nuclear ($\simgt2$~kpc)  X-ray sources at $z=0.03$--0.25 in  
late-type host galaxies within the two GOODS fields (i.e., the two Chandra Deep Fields).
The combination of the superb spatial resolution and
 great depth of the $HST$ ACS and $Chandra$ ACIS coverage in the GOODS fields is critical to the identification
of these sources.  We extend the study of this enigmatic population up to higher redshifts and larger look-back
 times than has been possible before, finding that the fraction of optically luminous galaxies exhibiting 
luminous off-nuclear sources at $z\approx0.1$ is larger than at the current time ($\approx36$\% versus $\approx8$\%).
The X-ray luminosities of the GOODS off-nuclear X-ray sources are comparable to or slightly greater than 
the integrated point-source X-ray luminosities of nearby galaxies, providing
a possible constraint on the X-ray Luminosity Function at higher star-formation rates.
X-ray variability demonstrates that several of these sources are 
likely single accreting black holes; for the majority, however, the number of X-ray counts is
too low for X-ray variability studies.   

\end{abstract}


\setcounter{footnote}{5}

\section{Introduction \label{intro}}

A small, but appreciable, fraction ($\simlt 6$\%) of the 0.5--8~keV cosmic X-ray background
arises from fairly normal and starburst galaxies rather than from
accretion onto supermassive black holes \citep[e.g.,][]{davoXI,HornOBXF}.
At $z\simlt0.25$  the X-ray emission of galaxies may be resolved with
$Chandra$, whose PSF corresponds to scales $\simlt 2.1$~kpc (0\farcs5).  Indeed, studies of the normal and
starburst galaxies in the $Chandra$ Deep Field (hereafter CDF) surveys have shown that $\simgt 20$\% of X-ray
detected galaxies at $z\simlt0.2$ appear to have X-ray emission originating outside their centers 
\citep[e.g., ][]{HornOBXF}.
These X-ray sources are the only detected X-ray emission within their host galaxies
 and have 0.5--8~keV luminosities of $\simgt 10^{39}$~\lumin , 
indicating that they are members of the off-nuclear ultraluminous X-ray source (ULX) population
that was discovered during the \einstein\ era \citep[e.g.,][]{Fabbiano89}.
ULX sources have X-ray luminosities typically exceeding that expected for
spherically symmetric Eddington-limited accretion onto ``stellar-mass"
(5--20~\msun) black holes.
They are thus ``ultraluminous" as compared
to other sub-galactic X-ray emitters (e.g., supernova remnants and X-ray binaries) but of
course much less luminous than most cosmic X-ray emitters (e.g.,AGN).
ULX sources may be powered by accretion onto stellar-mass black holes
\citep[e.g., the X-ray emission may be anisotropic; ][]{King01,Makishima00}.  They may also represent a class of   
intermediate-mass black holes \citep[$\approx 500$--1000~\msun, hereafter IMBH; e.g., ][]{Colbert99},
ultraluminous supernova remnants \citep[e.g.,][]{Blair01},  or the summed
emission from a collection of less luminous, unresolved X-ray sources.

 \chandra\ and \xmm\ studies of the X-ray colors and 
variability of ULX sources have shown that the majority are likely 
accreting sources rather than supernova remnants \citep[][]{Roberts02,Fabbiano03} and
further that the majority likely have stellar mass \citep[i.e., not IMBH; ][]{Zezas02}. 
The properties of stellar-mass ULX sources differ with galaxy type \citep[e.g.,][]{IrwinBinaries}
and with galaxy properties such as mass and starburst activity \citep[e.g.,][]{ColbertXLF2003}. 
It thus appears there are at least two different ULX formation mechanisms.
The formation mechanism for the 
few \citep[but confidently identified; e.g., ][]{MillerULX} IMBH candidates is less understood.

ULX sources are critically important in the study of X-ray emission from star-forming galaxies 
as they may dominate the point-source X-ray luminosity of e.g., spiral galaxies 
\citep[][]{ColbertXLF2003} . Despite the vast amount of information 
being assembled on these objects, there is still relatively little known about their 
prevalence in the field-galaxy population.   The very deep, high spatial resolution, $BVIz$ coverage
of the Great Observatories Origins Deep Survey \citep[GOODS;][]{MauroGOODSletter} 
is ideal for characterizing the off-nuclear X-ray sources as we may
resolve structure within the host galaxies and determine optical extent down
to lower surface brightness than would be possible with comparably deep ground-based data.

\section{Sample Construction}

The 0.5--8~keV flux of a $2\times10^{39}$~\lumin ~source at $z=0.1$ is $\approx6.8\times10^{-17}$~\flux, requiring the depth of the CDFs
\citep[][]{Giacconi02,davocatalog}; we use the X-ray catalogs 
of \cite{davocatalog} which have been constructed for both fields. 
The X-ray positions in these catalogs are good to
\hbox{0\farcs3--1\farcs0}; the positional accuracy depends on both the X-ray source's distance from the aim point (as the PSF size increases)
and on the number of detected counts.  
Over the central $\approx1$~arcmin$^{2}$, the best 0.5--2~keV S/N$=$3 sensitivity achieved is $\approx2.5\times10^{-17}$~\flux 
~in the North and $\approx 5.2\times10^{-17}$~\flux in the South, dropping
by a factor of $\approx1.8$--6 in the north and $\approx1.2$--10 in the south
at the edges of the GOODS fields.

The GOODS ACS photometric data are complete to $z_{\rm AB}\approx24$--25, depending on the half-light radius of the
galaxy \citep[see Figure~4 of][]{MauroGOODSletter}.  
In the \hbox{CDF-N}, the main sources of optical spectroscopic data used were  
\cite{Cohen00} and \cite{BargerCatalog2002}.  In the \hbox{CDF-S}, 
we used the Two Degree Field (2dF) survey redshift catalog 
for the three optically brightest galaxies \citep{Colless01} and
spectroscopic redshifts of $z<0.3$ galaxies from the CDF-S team
(G. Szokoly et~al., in preparation and G. Szokoly 2003, private communication).  
When there was no spectroscopic redshift, we used the
photometric redshifts of B. Mobasher et~al., in preparation, computed 
from both the ACS data and ground-based near-infrared and 
optical data.

As discussed in \cite{davocatalog}
and \cite{HornOBXF}, matching radii slightly larger than the X-ray positional accuracy
are acceptable for optically bright galaxies ($V\simlt21$~mag ) as the effective
significance of the match is ``boosted" by the relative rarity of these galaxies
on the sky.  We additionally require that the host galaxies have $z<0.25$ so that the
median $Chandra$ astrometric accuracy of 0\farcs6 corresponds to $<2.5$~kpc; this redshift
requirement further boosts the significance of the matches over merely requiring $V<21$~mag.  
Based on the optical ($V<21$~mag,$z<0.25$) and X-ray source densities, we expect $\approx0.3$ 
false matches per field when using a 2\farcs0 matching radius.  
We have also searched at larger offsets ($2$\farcs0--3\farcs0) for sources with $V<20$~mag and $z<0.25$, this raises the 
possible number of false matches to a total of 0.7.  We have explored extending the search to $V<22$~mag 
but find no further off-nuclear sources at $z < 0.25$.  Finally, we include one source 
with $V=16.7$~mag and offset $\approx$4\farcs7.  

The X-ray/optical offsets must exceed the X-ray positional errors \citep[80--90\% confidence; see][]{davocatalog}
by at least 40\% to be considered
off-nuclear.  Sources with offsets of 0\farcs8--1\farcs0 are just marginally off-nuclear,
 and it may be that slight X-ray extent is creating the offset.
We find 10 off-nuclear X-ray sources at $z\approx0.02$--0.25 matched with the 
main catalogs of \cite{davocatalog}, with the majority (7/10) in the northern field.  
The X-ray source and host-galaxy  properties are listed in Table~1; the coordinates in the 
table and throughout this Letter are X-ray coordinates.  Spectroscopic redshifts were 
available for 9 of the 10 sources (median $z=0.109$).  The sources are indeed faint, with 
a median number of 0.5--8~keV counts of 28.7.  Using the sensitivity maps derived in
\cite{davocatalog}, we have determined that only three of the northern ULX sources 
would have been clearly detected in the southern field  had they been placed at the same 
off-axis locations.  This indicates that the chief difference between the northern and 
southern samples is the X-ray sensitivity, not cosmic variance.

\section{Properties of the Off-Nuclear X-ray Soruces}
\noindent

\subsection{Host Galaxy Properties}
\noindent

The $HST$ $V$-band images of all the ULX host galaxies, which range in optical brightness 
from $V=16.7$--20.6~mag,  are shown in Figure~1.  
The median absolute magnitude of the host galaxies is $M_{V}=-20.1$, and the least luminous galaxy 
has $M_{\rm V}=-18.2$ .

From visual inspection it is clear that these galaxies are primarily late-type/spiral galaxies.
However, there are no early-type galaxies at $z<0.25$ in the GOODS fields, so this apparent
bias toward late-type/spiral galaxies is due to the limited volume sampled by the two GOODS fields.
Characterizing their morphologies using the CAS morphological
system of \cite{cc03},  the ULX host galaxies are found to be fairly representative of the
morphological types of field galaxies in the GOODS area, with no particularly strong
bias toward large asymmetries.

The GOODS off-nuclear X-ray sources are fairly evenly split among those associated 
with a resolved optical source or other optical structure and those
which  appear associated based on statistical likelihood alone.  Among the latter, 
there is one galaxy for which the ULX is just marginally offset \hbox{($\approx0$\farcs9; 
 J123715.9+621158)} and there is a fairly bright optical point source near the
galactic nucleus. 

\subsection{X-ray Properties}
\noindent

The 0.5--2~keV luminosities of the ULX sources range from $9\times 10^{38}$~\lumin to a few times $10^{40}$~\lumin.
Note that there have been local ULX sources observed at X-ray luminosities as high as $10^{40}$~\lumin
\citep[e.g., the strongly variable ULX in NGC~3628; ][]{StricklandULX}.  
Some spectral constraints may be placed using the various X-ray bands of \cite{davocatalog}.
Only one source is constrained to be X-ray hard (J033234.7$-$275533, $\Gamma < 0.17$). There is just one 
ultrasoft source detected {\it only} in the 0.5--1~keV ``ultrasoft" band (J033220.3$-$274555; see footnote in Table~1).
Although poorly constrained, its soft X-ray band ratio is consistent with $kT<0.6$~keV thermal bremsstrahlung emission
(using its photometric redshift $z=0.23$).  In general, the 1--2~keV data demonstrate that the ULX sources 
do not have the ultrasoft signatures of the cooler accretion disks expected around intermediate-mass black holes. 
Also, they demonstrate that  the X-ray emission does not arise from diffuse hot gas in star-forming regions--
for instance in M101, which is dominated by its hot ISM, the highest surface brightness regions emit
very little above 1 keV \citep[see Figure~4 of ][]{Kuntz03}.
Stacking the X-ray data for all the objects that are not individually detected in the hard band, 
we find that the average spectral slope of these sources is $\Gamma > 1.5$.  

For only the brightest few X-ray sources are there sufficient numbers of photons to perform variability 
analysis.   Following the methods of M. Paolillo et~al., in preparation,
we find that five of the off-nuclear sources are likely variable on timescales of days to months
with detected variation amplitudes of $\approx2$--3.  The most notable instance of variability is
the X-ray hard source J033234.7$-$275533 (98\% probability in the 2--8~keV band).

\section{Discussion}

Based on the sky density of $z<0.25$,$V<21$ field galaxies within the Caltech Faint Galaxy Redshift 
Survey \citep{Cohen00},  we have detected off-nuclear X-ray emission from $\approx36$\% of field 
galaxies at these redshifts.   For comparison with the nearby Universe, we consider the complete, 
volume-limited Palomar optical spectroscopic survey sample of \cite{Ho97}. 
\cite{Ho01} chose galaxies from this survey for $Chandra$ follow-up if they had optical evidence for LLAGN activity;  otherwise this 
is one of the currently least-biased sample of X-ray observed galaxies in the nearby Universe.  The optical brightness of the \cite{Ho01}
sample is well-matched to GOODS as it was chosen to be fairly optically bright ($B_{T} < 12.5$~mag).  Based on the
$V$-band flux density information in the NASA/IPAC Extragalactic Database (NED), 36 of the 41
galaxies in the \cite{Ho01} sample would have $V<21$ at $z<0.15$.  \cite{SipiorTHESIS} has compiled a catalog of X-ray 
point sources for these 36 galaxies; we use this catalog to determine that $8^{+8}_{-5}$\% of galaxies in the 
nearby Universe have point sources with 0.5--8~keV luminosities $>2\times10^{39}$~\lumin .  The GOODS off-nuclear
source fraction is higher at $36^{+24}_{-15}$\% and even allowing for false matches is higher at 
$32$\% .  Note that even $Chandra$ is limited by spatial resolution, so
this off-nuclear source fraction may be considered as a {\it lower} limit, the real fraction may in fact be higher.
This indicates that there may be a higher prevalence of these more luminous ULX sources at $z\approx0.1$.

\cite{ColbertXLF2003} note that the point-source X-ray luminosities of galaxies, $L_{\rm XP}$ (0.3--8~keV),
 are generally dominated by the most luminous X-ray binaries in galaxies, finding a correlation with  
$K$-band luminosity: $L_{K} \propto L_{\rm XP}^{+0.97}$.   While there is an apparent correlation 
between the GOODS off-nuclear 0.5--8~keV X-ray luminosity and $L_{K}$ (calculated as $\nu f_{\nu}$; see 
Table~1 and Figure~2), we note that the most luminous object is also at the
highest redshift so this is likely a ``distance" effect.  The correlation of the X-ray properties 
with the host-galaxy properties does indicate that the off-nuclear sources are not chance
optically faint background AGN \citep{davofaint}.  

For comparable host-galaxy $K$-band luminosities, the X-ray luminosities of the
GOODS off-nuclear X-ray sources are on average $\approx2.2^{+0.6}_{-0.5}$ times higher
than the expected total galaxy X-ray point-source luminosity \citep[we converted $L_{\rm XP}$ from 
0.3--8~keV to 0.5--8~keV, assuming $\Gamma=2$; ][]{ColbertXLF2003}.   
The implication is that there is a global difference between the $z\approx0.1$ GOODS sample and the local Universe. 
We note that even if there is some confusion in our sample, with a few luminous point sources
contributing from one star-forming region, the result that the X-ray luminosities of the GOODS off-nuclear sources
exceed the {\it total} expected galaxy point-source X-ray luminosity is robust.

The maximum luminosity for point sources within a galaxy
can be determined from the galaxy's X-ray luminosity function; it is simply the luminosity, $L$, for which
the cumulative X-ray Luminosity Function (XLF), $N(>L_{max})\approx1$.   $L_{max}$ and the XLF are 
dependent upon both galaxy mass and SFR.  The $K$-band
luminosity is a fairly good measure of galaxy mass, so this indicates the difference between the two samples is due to differences
in SFR, i.e., the globally elevated SFR at $z\approx0.1$.  

Constraining the high-luminosity/high-SFR end of the XLF is difficult due to the typically 
low number of ULX sources per galaxy \citep[the one exception in the nearby Universe is the pair 
of Antennae galaxies; ][]{Zezas02}.  
Based on measurements of both the infrared and $B$-band luminosity densities of galaxies, the SFR
of the Universe was $\approx60$\% higher at $z\approx0.1$ \citep{Treyer98}.  The difference in X-ray luminosities
of a factor of $\approx2.2$, assumed to be due to the difference in SFR,
 thus implies that $L_{max}, L_{\rm XP} \propto $SFR$^{2.3}$, corresponding to an XLF slope of $\approx 0.4$.
For comparison, the Antennae XLF has slope $\alpha=0.47$; our number is thus in coarse agreement with the
extension of the Antennae XLF up to higher X-ray luminosities.
 While detailed measurements of the GOODS host-galaxy SFRs and masses 
is beyond the scope of this letter, this indicates that future such measurements (possible 
with $SIRTF$) may provide a very useful constraint on $L_{max}$ and the XLF of galaxies.

\acknowledgments

We gratefully acknowledge the financial support of
$Chandra$ fellowship grant PF2-30021 (AEH),
\chandra\ X-ray Center grant G02-3187A (FEB,WNB), and 
NSF CAREER award AST-9983783 (DMA,FEB,WNB).
DMA also gratefully acknowledges the generous support from the Royal Society.
Additional support was also provided by NASA through grant GO09583.01-96A from the 
Space Telescope Science Institute, which is operated by the Association of 
Universities for Research in Astronomy, under NASA contract NAS5-26555.   
RC acknowledges suport from the $SIRTF$ Legacy Science Program, provided through an award
issued by the Jet Propulsion Laboratory (JPL), California Institute of Technology (Caltech) under NASA 
contract 1407.
We thank Edward Colbert for reading a draft of the manuscript and
Edward Colbert, Gyula Szokoly and Piero Rosati for sharing data.  
 This research has made use of the NASA/IPAC Extragalactic Database which is 
operated by JPL under contract with NASA.
%



\clearpage
\noindent
\begin{deluxetable}{rcrrrcccrrrc}
\tabletypesize{\tiny}
\tablewidth{0pt}
\tablecaption {Candidate Off-Nuclear X-ray Sources in the GOODS fields}
%
%
\tablehead{
\multicolumn{1}{c}{Coordinates$^{\rm a,b}$ } 				&
\multicolumn{1}{c}{Pos. Err/$\Delta$Opt$^{\rm b}$}			&
\colhead{$\Gamma^{\rm a}$}                                              &
\multicolumn{2}{c}{$f_{\rm X}$ ($10^{-16}$~cgs)}      &
\colhead{$V$} &
\multicolumn{1}{c}{$z$$^{\rm d}$}                                                 &
\multicolumn{1}{c}{$z$}							 &
\multicolumn{1}{c}{Offset}                                              &
\multicolumn{1}{c}{$\log L_{\rm HB}$}					&
\multicolumn{1}{c}{$\log L_{\rm SB}$}                                   &
\multicolumn{1}{c}{$\log L_{\rm B}$}
\\
\multicolumn{1}{c}{(J2000)}                                     	&
\multicolumn{1}{c}{($^{\prime \prime}$)/($^{\prime \prime}$)}		&
\colhead{}								&
\colhead{HB}                                      			&
\colhead{SB}                                      			&
\colhead{mag$^{\rm c}$}                                                &
\colhead{}                                                      	&
\colhead{src$^{\rm d}$}						&
\multicolumn{1}{c}{(kpc)}                                       	&
\multicolumn{3}{c}{(\lumin)$^{\rm e}$}						
}
\startdata
$033220.35-274555.3 $ &  0.6 / $  0.8$  & $ \sim $ & $<  1.23$ & $<  1.65$ & 20.6 & $ 0.23^{+ 0.16}_{- 0.16}$   &P  &3.7 & $< 40.7$   & $ 39.9^{\rm f}$   &  43.4    	 \\  
$033230.01-274404.0 $ &  0.6 / $  0.8$  & 1.87         & 12.00     & 12.00     & 17.4 & $ 0.075 $  &1  &1.9 &  40.1      &  40.0      &  43.4    	 \\  
$033234.73-275533.8 $ &  0.9 / $  4.8$  & $<0.17$      & 16.90     & $< 16.90$ & 16.7 & $ 0.038 $  &1  &3.9 &  39.8      & $< 38.7$   &  42.9    	 \\  
$ 123637.18+621135.3 $ &  0.6 / $  2.2$  & $ \sim $ & $<  1.07$ &  1.43     & 18.3 & $ 0.078 $  &3  &3.6 & $< 39.4$   &  38.9      &  43.2    	 \\  
$ 123641.81+621132.4 $ &  0.6 / $  1.2$  & $>1.73$      & $<  2.31$ &  2.31     & 20.0 & $ 0.089 $  &3  &2.2 & $< 39.5$   &  39.4      &  42.7    	 \\  
$ 123701.49+621846.1 $ &  0.6 / $  3.2$  & $>1.56$      & $<  3.15$ &  3.15     & 19.0 & $ 0.232 $  &2  &12.0 & $< 40.6$   &  40.3      &  43.9    	 \\  
$ 123701.98+621122.3 $ &  0.6 / $  0.9$  & $ \sim $ & $<  1.01$ &  1.35     & 19.4 & $ 0.136 $  &3  &3.0 & $< 39.9$   &  39.4      &  43.3    	 \\  
$ 123715.94+621158.5 $ &  0.6 / $  0.8$  & $ \sim $ & $<  1.06$ &  1.42     & 18.7 & $ 0.112 $  &3  &2.3 & $< 39.8$   &  39.3      &  43.1    	 \\  
$ 123721.60+621247.0 $ &  0.6 / $  2.4$  & $ \sim $ & $<  1.81$ &  2.43     & 18.0 & $ 0.106 $  &3  &5.2 & $< 39.9$   &  39.3      &  43.5    	 \\  
$ 123723.44+621048.2 $ &  0.7 / $  3.0$  & $ \sim $ & $<  1.40$ &  1.88     & 19.5 & $ 0.113 $  &2  &6.1 & $< 40.0$   &  39.3      &  43.0    	 \\  
\enddata
\tablenotetext{a}{X-ray coordinates, photon indices ($\Gamma$), and fluxes are from \cite{davocatalog}.  For $\Gamma$, a
$\sim$ symbol indicates that the photon index is not well-constrained due to low number of counts, $\Gamma=1.4$ is the
value assumed in the calculation of the X-ray fluxes. }
\tablenotetext{b}{The X-ray coordinates and the offset between the X-ray and optical source has been corrected following the methods of
Koekemoer et~al., in preparation and Bauer et~al., in preparation.}
\tablenotetext{c}{$HST$ $ACS$ $V$ (F606W) filter magnitudes of S. Ravindranath et al., in preparation.}
\tablenotetext{d}{``P" indicates that the redshift is photometrically determined (B. Mobasher et al., in preparation).
For photometric redshifts we quote the 90\% confidence range.  The sources for spectroscopic redshifts (which have very
small errors, $<< 0.01$) are as follows: (1): Two Degree Field (2dF) Redshift Survey  \cite{Colless01}, (2):  CDF-N 1~Ms spectroscopic catalog
\cite{BargerCatalog2002} and (3): Caltech Faint Galaxy Redshift Survey \citep[CFGRS; ][]{Cohen00}}
\tablenotetext{e}{All luminosities are calculated using the same cosmology as \cite{MauroGOODSletter}.}
\tablenotetext{f}{033220.3-274555 was only detected in the 0.5--1~keV band.  Its soft band X-ray luminosity is quoted in that band.}

\end{deluxetable}


\clearpage



\begin{figure}[t!]
\figurenum{1}
\centerline{\includegraphics[scale=0.80,angle=0]{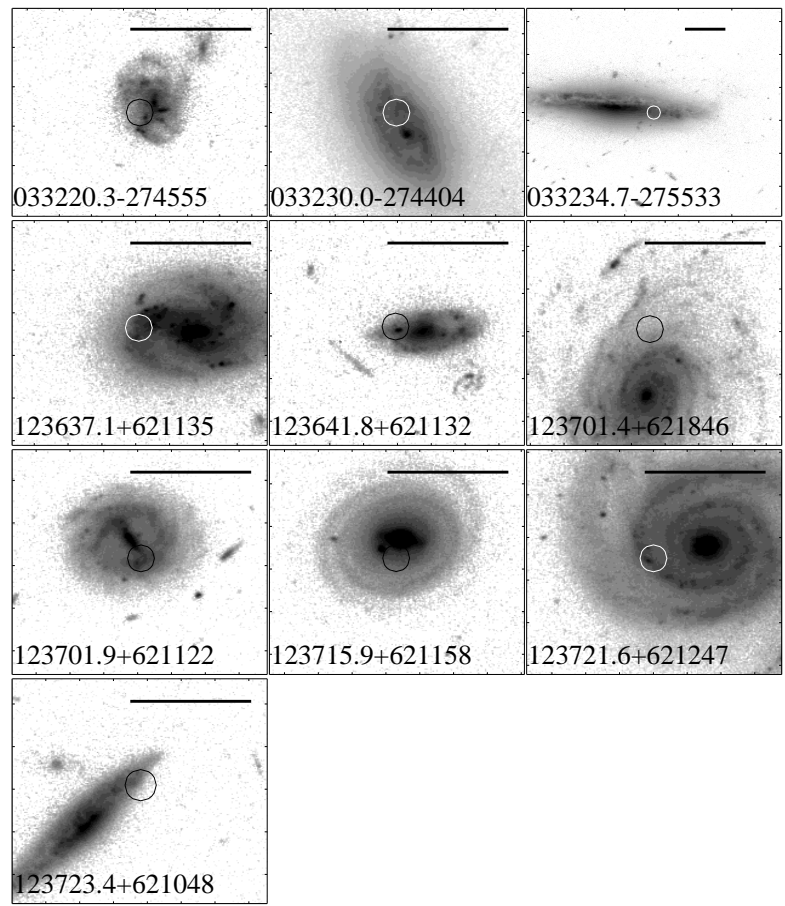}}
\caption[]{
The $HST$ ACS $V$-band thumbnail images of the ULX
source host galaxies in the two GOODS fields.   
The line in the upper right corner indicates 5\farcs0 for scale.
The circles indicate the X-ray positional error.}

\end{figure}

\begin{figure}[t!]
\figurenum{2}
\centerline{\includegraphics[scale=0.70,angle=0]{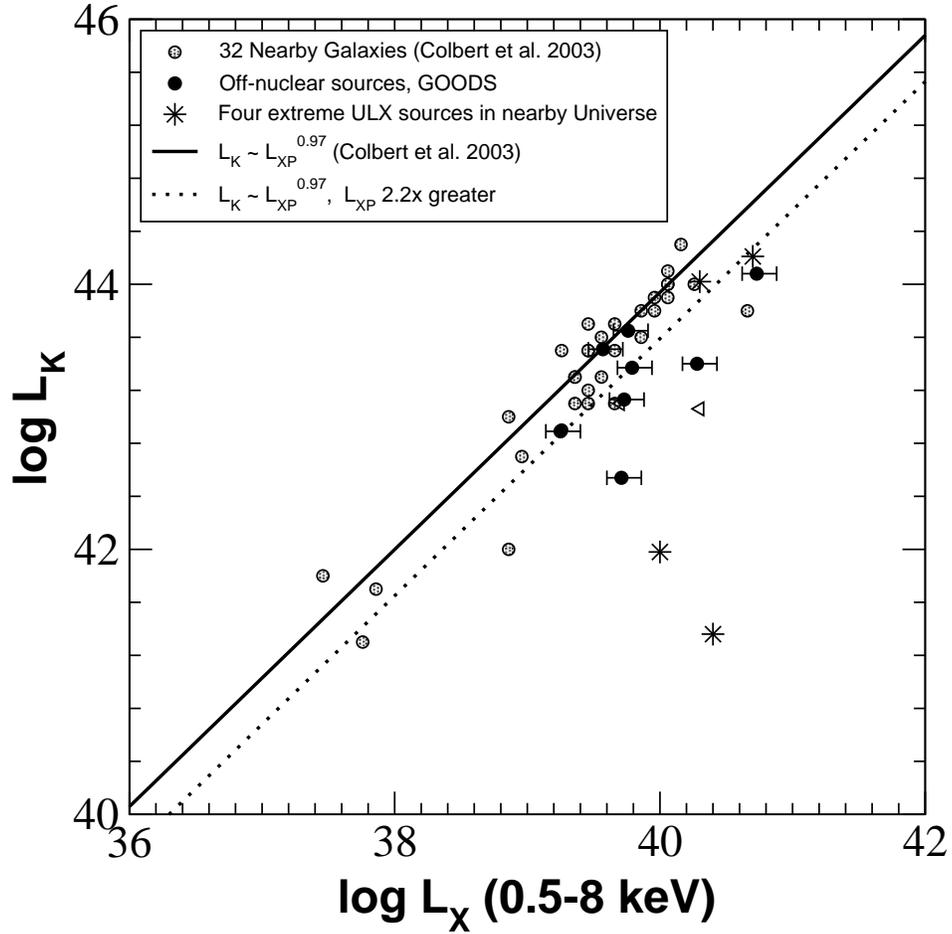}}
\caption[]{$K$-band luminosity of the host galaxies versus the off-nuclear X-ray source
luminosity.  The filled circles are the data from this paper, the two triangles indicate upper limits.
The open circles indicate data on 32 nearby galaxies from Colbert et al. (2003); the X-ray luminosity
is the {\it total} X-ray point source luminosity for these galaxies.  The 0.3--8~keV X-ray luminosities
of Colbert et al. (2003) have been converted to 0.5--8~keV assuming $\Gamma=2$.  The star symbols indicate
the ULX source X-ray luminosities and host galaxy $K$-band luminosities for four extreme instances of
ULX sources in the nearby Universe: NGC~1313, Holmberg II, NGC~4565 and NGC~4254. The X-ray data are from
\cite{Colbert2002}.
}

\end{figure}

\end{document}